\documentclass[a4paper,preprintnumbers,showpacs,twocolumn,superscriptaddress,nofootinbib,amsmath,amssymb]{revtex4-1}
\usepackage{color,hyperref}
\usepackage{times}
\usepackage{mathrsfs}
\hypersetup{colorlinks=true,linkcolor=blue,citecolor=magenta,filecolor=magenta,urlcolor=blue}

\usepackage{enumitem}
\usepackage{graphicx,subfigure}
\usepackage{dcolumn}
\usepackage{bm}
\usepackage{color}

\def\beq{\begin{equation}}
\def\eeq{\end{equation}}
\def\bear{\begin{eqnarray}}
\def\ear{\end{eqnarray}}
\def\nn{\nonumber}

\begin{document}

\title[Tidal forces in parametrized spacetime: RZ  parametrization]{Tidal forces in parametrized spacetime: Rezzolla-Zhidenko parametrization}

\author{Bobir Toshmatov}
\email{toshmatov@astrin.uz} 
\affiliation{New Uzbekistan University, Mustaqillik Ave. 54, Tashkent 100007, Uzbekistan}
\affiliation{Ulugh Beg Astronomical Institute, Astronomy str. 33, Tashkent 100052, Uzbekistan}
\affiliation{Research Centre of Theoretical Physics and Astrophysics, Institute of Physics, Silesian University in Opava, Bezru\v{c}ovo n\'{a}m. 13, CZ-74601 Opava, Czech Republic}

\author{Bobomurat Ahmedov}
\email{ahmedov@astrin.uz} 
\affiliation{Ulugh Beg Astronomical Institute, Astronomy str. 33, Tashkent 100052, Uzbekistan}
\affiliation{Institute of Fundamental and Applied Research, ``TIIAME" National\\ Research University, Kori Niyoziy 39, Tashkent 100000, Uzbekistan}
\affiliation{National University of Uzbekistan, Tashkent 100174, Uzbekistan}

\begin{abstract}
We investigate the tidal forces exerted by a spherically symmetric static parametrized black hole. Our analysis reveals that the radial and angular components of the tidal forces exerted by the black hole can exhibit both positive and negative values near the black hole, depending on matters of the spacetime parameters. Unlike the scenario with the Schwarzschild black hole, where the radial tidal force (angular tidal force) is always stretching (compressing) and becomes infinite at the center of the spacetime, the parametrized black hole allows for finite and compressing (stretching) forces within the event horizon. Additionally, we derive the geodesic deviation equations for a particle in free fall and proceed to solve them through numerical methods. Our analysis demonstrates that the spacetime parameters $\epsilon$ and $a_1$ exhibit contrasting influences on the magnitudes of the physical quantities associated with tidal effects.
\end{abstract}

\maketitle

\section{Introduction}

The mysterious and fascinating characteristics of black holes have consistently captivated both physicists and astrophysicists. The black holes are characterized by their strong gravitational attraction and special boundaries that distort space so much that even light cannot escape from them. One of the most intriguing aspects of this gravitational dominance is the phenomenon of tidal forces, which emerge when an external gravitational field acts upon an extended body, resulting in the deformation and stretching (shrinking) of its constituent parts. Tidal effects around black holes have attracted significant attention due to their profound implications across various astrophysical contexts \cite{MTW,Carroll:2004st,Poisson:2009pwt,2006gere:Hobson}. From inspirals of compact objects into black holes to the interactions within stellar binaries and the behavior of matter in active galactic nuclei, tidal forces play a pivotal role in shaping the dynamics and observable phenomena in these extreme gravitational environments \cite{1988Natur523R,Janis:1968zz,Flanagan:2007ix,Kesden:2011ee,LIGOScientific:2019eut,Strubbe:2009qs,Haas:2012bk,Gair:2012nm,Wang:2020xwn,Kumar:2016zlj,Kesden:2011ee,Williams:2022vct}. It is well known that in the Schwarzschild spacetime, a freely falling particle experiences a consistent stretching in the radial direction and compression in the angular direction during its trajectory toward the spacetime center \cite{Hong:2020bdb}. However, distinctive behaviors of tidal effects emerge in the context of Reissner-Nordstr\"{o}m and dirty black holes, showcasing a notable transition from radial stretching to compressing, and from angular compressing to stretching, occurring between the inner and outer horizons of the spacetime \cite{Crispino:2016pnv,Hong:2020bdb,LimaJunior:2022gko,LimaJunior:2020fhs,Liu:CPC:2022,Lima:2020wcb}. Furthermore, various phenomena associated with tidal effects encompassing diverse spacetimes in the realm of gravity theories have been explored in previous studies \cite{Magorrian:1999vm,Wang:2003ny,Evans:1989qe,Esquej:2006ff,Ishii:2005xq,Cheng:2013uva,Goel:2015zva,Shahzad:2017vwi}.

Despite Einstein's general theory of relativity having proven highly successful in explaining gravity's behavior in many scenarios, there are still some unresolved problems and observational data that require potential limitations or modifications to this theory. As a result, researchers have developed various alternative theories of gravity to address these discrepancies, to explore unexplained phenomena such as dark matter and dark energy, and seek a more comprehensive understanding of the universe's fundamental forces. Parametrizing black hole spacetimes is crucial for theoretical exploration, comparative analysis, testing gravity theories, interpreting gravitational wave signals, modeling astrophysical phenomena, and facilitating education. It allows scientists to investigate the effects of different parameters on black hole geometry, enhancing our understanding of these objects and their role in the universe. In recent years a numbers of parametrically deformed metrics have been proposed, such as the Johannsen-Psaltis metrics \cite{Johannsen:2011dh,JP-non-Kerr}, their extensions \cite{PhysRevD.89.064007,Carson:2020dez}, and the Rezzolla-Zhidenko metrics \cite{Rezzolla:2014mua,Konoplya:2016jvv}. Since then, researchers have diligently sought to find constraints on the parameters of the parametrized spacetime through analysis of various observational data and the exploration of relevant physical phenomena \cite{Medeiros:2019cde,Psaltis:2020ctj,EventHorizonTelescope:2021dqv,Lara:2021zth,Tian:2019yhn,AA:PRD:2021,Nampalliwar:2019iti,DeLaurentis:2017dny,Toshmatov:2021hdy,Tripathi:2021rqs}.

Our study in this paper focuses on the tidal forces produced by spherically symmetric static parametrized black holes proposed by Rezzolla and Zhidenko \cite{Rezzolla:2014mua}, encompassing exploration of both radial and angular components of these forces. In the following sections, we will discuss our methodology, including the geodesic deviation equations for particles in free fall and the numerical techniques employed to analyze their behavior in the presence of tidal forces. We will also delve into the intricate relationship between spacetime parameters and the physical quantities linked to tidal effects, revealing how variations in these parameters can lead to divergent behaviors. Hence, in Section \ref{sec-Eqs}, we provide a concise overview of the parametrized spacetime concept, while in Section \ref{sec-rad-geod}, we delve into the investigation of radial geodesics concerning massive test particles. The equation for geodesic deviation within the parametrized spacetime framework is introduced in Section \ref{sec-geod-dev}. Furthermore, Sections \ref{sec-rtf} and \ref{sec-atf} are dedicated to the comprehensive examination of radial and angular tidal forces, along with the numerical solutions of the geodesic deviation equations within the parametrized black hole spacetime, respectively. To conclude, our key findings are summarized in Section \ref{sec-conc}. Throughout the paper, we adopt the metric signature ($-$,$+$,$+$,$+$) and set the speed of light and the Newtonian gravitational constant equal to the unity, $c=G=1$.

\section{Basic equations}\label{sec-Eqs}

As indicated by Rezzolla and Zhidenko in their work \cite{Rezzolla:2014mua}, the line element characterizing spherically symmetric and static black holes can be elegantly expressed in a parametrized form as
\begin{equation}\label{line-element}
ds^2=-N^2(r)dt^2+\frac{B^2(r)}{N^2(r)}dr^2+r^2 d\Omega^2\ ,
\end{equation}
where the metric function $N^2(r)$ depends on the radial coordinate $r$ only and the term $d\Omega^2 \equiv d\theta^2+\sin^2\theta d\phi^2$ denotes the solid angle. By introducing a new dimensionless variable, expressed in terms of the event horizon radius $r_0$, it becomes feasible to streamline the radial coordinate as follows:
\begin{equation}
x\equiv 1-\frac{r_0}{r}\ ,
\end{equation}
in terms of the new coordinate, the event horizon corresponds to the value of $x=0$, while spatial infinity is represented at $x=1$, effectively establishing a range of $x\in[0,1]$. Furthermore, we assume that the metric function is given as
\begin{equation}\label{N2}
N^2(r)=x A(x)\ . 
\end{equation}
The functions $A$ and $B$ in (\ref{N2}) are given in terms of the bumpy parameters as \cite{Rezzolla:2014mua}
\begin{eqnarray}\label{asympfix_1}
A(x)&=&1-\epsilon (1-x)+(a_0-\epsilon)(1-x)^2+{\tilde A}(x)(1-x)^3\ ,\nonumber\\
B(x)&=&1+b_0(1-x)+{\tilde B}(x)(1-x)^2\ .
\end{eqnarray}
In expressions (\ref{asympfix_1}), the parameters $\epsilon$, $a_0$, and $b_0$ serve as new coefficients, while ${\tilde A}$ and ${\tilde B}$ are introduced to characterize the metric properties at the event horizon and spatial infinity through the Pad\'e series as \cite{Rezzolla:2014mua}
\begin{eqnarray}\label{contfrac_1}
{\tilde A}(x)&=&\frac{a_1}{\displaystyle 1+\frac{\displaystyle
    a_2x}{\displaystyle 1+\frac{\displaystyle a_3x}{\displaystyle
      1+\ldots}}}\ ,\nonumber\\
{\tilde B}(x)&=&\frac{b_1}{\displaystyle 1+\frac{\displaystyle
    b_2x}{\displaystyle 1+\frac{\displaystyle b_3x}{\displaystyle
      1+\ldots}}}\ ,\nonumber
\end{eqnarray}
where $a_1, a_2, a_3,\ldots$ and $b_1, b_2, b_3,\ldots$ are dimensionless constants. Notably, the dimensionless coefficient $\epsilon$ holds a direct relationship with the event horizon via
\begin{equation}
\epsilon=-\left(1-\frac{2M}{r_0}\right)\ ,
\end{equation}
where $M$ is the ADM mass. $\epsilon$ measures the deviations of event horizon $r_0$ from $2M$ and it is an important parameter because one can recast all the other coefficients in terms of it.

The metric functions $N^2(r)$ and $B(r)$ can be written explicitly in terms of the expansion parameters as
\begin{eqnarray}\label{fr}
&&N^2(r)=\nonumber\\
&&\left( 1-\frac{r_0}{r}\right) \left[1-\epsilon\frac{r_0}{r}+(a_0-\epsilon)\left(\frac{r_0}{r}\right)^2+a_1\left(\frac{r_0}{r}\right)^3...\right]\ ,\nonumber\\
&&B(r)=1+b_0\frac{r_0}{r}+b_1\left(\frac{r_0}{r}\right)^2+...\ . 
 \end{eqnarray}
If $\epsilon=0$, $a_i=0$, $b_i=0$ with $i=0,1,2,\ldots$, one recovers the standard spherically symmetric Schwarzschild spacetime. Furthermore, the requirement for consistency with general relativity at the 1PN order has demonstrated that $a_0=0$ and $b_0=0$ \cite{Rezzolla:2014mua}. To simplify our subsequent calculations, we limit our consideration to the third order of $x$ terms, resulting in the following expressions for the metric functions:
\begin{eqnarray}\label{metric-functions}
&&N^2(r)=\left( 1-\frac{r_0}{r}\right) \left[1-\epsilon\frac{r_0}{r}\left(1+\frac{r_0}{r}\right)+a_1\left(\frac{r_0}{r}\right)^3\right]\ ,\nonumber\\
&&B(r)=1+b_1\left(\frac{r_0}{r}\right)^2\ .
\end{eqnarray}
With the groundwork established for the background spacetime, we can now proceed to dive into the study of the equations of motion for the test particle. To investigate the motion of both massive and massless particles using the same equations of motion, one can achieve this by setting the mass of the particle to zero in the latter scenario, effectively reverting to the former case. Notably, the symmetry of the spacetime metric (\ref{line-element}) gives rise to conserved momenta associated with the time and azimuthal coordinates, attributed to the spacetime's stationarity and spherical symmetry. These conserved momenta are termed the energy ($E$) and angular momentum ($L$) of the particle, respectively, as
\bear
&&N^2(r)u^t=E\ ,\label{ut}\\
&&r^2u^\phi=L\ .\label{up}
\ear
Henceforth, we adopt the terms specific energy and specific angular momentum instead of the energy and angular momentum of the massive particle with mass $m$, represented as $E\rightarrow E/m$ and $L\rightarrow L/m$, respectively. These concepts are defined on a per-unit-mass basis. Furthermore, to simplify matters, we restrict our analysis to particle motion confined to the equatorial plane. By applying the normalization condition $u^\mu u_\mu=-\varepsilon$, the momentum corresponding to the radial coordinate of the particle can be derived as:
\bear\label{ur}
B^2\left(u^r\right)^2=E^2-V_{\rm eff}, \quad V_{\rm eff}=N^2(r)\left(\varepsilon+\frac{L^2}{r^2}\right)\ ,
\ear
where $\varepsilon=0, 1$ for the massless and massive particles, respectively. The effective potential can be separated into three components with the first being of general relativity and the second and third being of the additional deviation terms from the general relativity on account of the parameters $\epsilon$ and $a_1$, respectively, as
\bear
V_{\rm eff}=V_{\rm eff}^{\rm GR}+\epsilon \delta V_{\rm eff1}+a_1 \delta V_{\rm eff2},
\ear
where
\bear
&&V_{\rm eff}^{\rm GR}=\left(1-\frac{r_0}{r}\right) \left(\varepsilon+\frac{L^2}{r^2}\right),\nonumber\\
&&\delta V_{\rm eff1}=-\frac{r_0}{r}\left(1-\frac{r_0}{r}\right)\left(1+\frac{r_0}{r}\right) \left(\varepsilon+\frac{L^2}{r^2}\right),\\
&&\delta V_{\rm eff2}=\left(\frac{r_0}{r}\right)^3\left(1-\frac{r_0}{r}\right) \left(\varepsilon+\frac{L^2}{r^2}\right).\nonumber
\ear
Thus, the equations of motion of the particle confined at the equatorial plane of the spherically symmetric parametrized spacetime is given by the followings:
\bear
&&u^t=\frac{E}{N^2}\ ,\nn\\
&&B^2\left(u^r\right)^2=E^2-V_{\rm eff}\ ,\nn\\
&&u^\phi=\frac{L^2}{r^2}\ .
\ear
It is well known that the particle moving along the circular orbit around the black hole has zero radial velocity and acceleration. 
\begin{figure}[th]
\centering\includegraphics[width=0.50\textwidth]{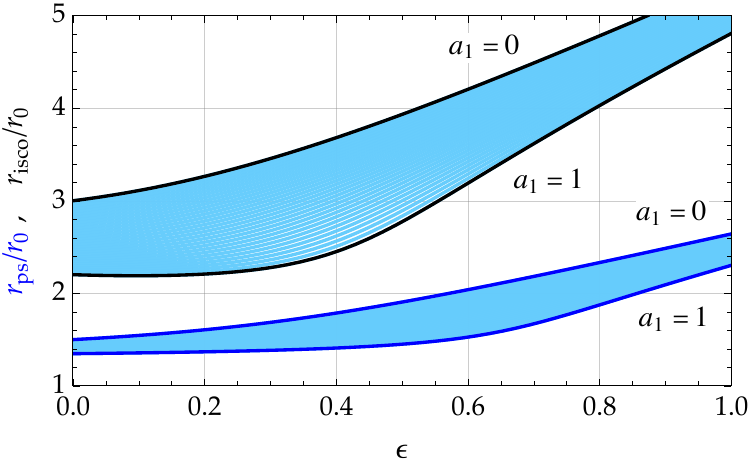}
\caption{Radii of photon sphere, $r_{\rm ps}$ (region between blue curves) and the innermost stable circular orbits, $r_{\rm isco}$ (region between black curves) as a function of the spacetime parameters. Where we set $a_1\in[0,1]$ with 0.01 steps.}\label{fig-orbits}
\end{figure}
Therefore, one can see from the above equations of motion that the radii of the characteristic circular orbits around parametrized black holes are independent from the parameters $b_i$. Despite, further computations concerning the circular motion of the test particle around parametrized black hole have been conducted in  \cite{Toshmatov:2021hdy,Toshmatov:2022kim}, in Fig. \ref{fig-orbits} we present the effect of the spacetime parameters $\epsilon$ and $a_1$ on the radii of the circular null geodesics (photon sphere) and the innermost stable circular orbit (ISCO). One can see from Fig. \ref{fig-orbits} that an observable trend emerges where the radii of circular orbits expand as the spacetime parameter $\epsilon$ increases. Conversely, the parameter $a_1$ brings about a reduction in the radii of these orbits. In essence, the parameter $\epsilon$ enhances the gravitational attraction, while the parameter $a_1$ weakens it.

\section{Radial geodesics of massive particle}\label{sec-rad-geod}

\begin{figure}[th]
\centering\includegraphics[width=0.50\textwidth]{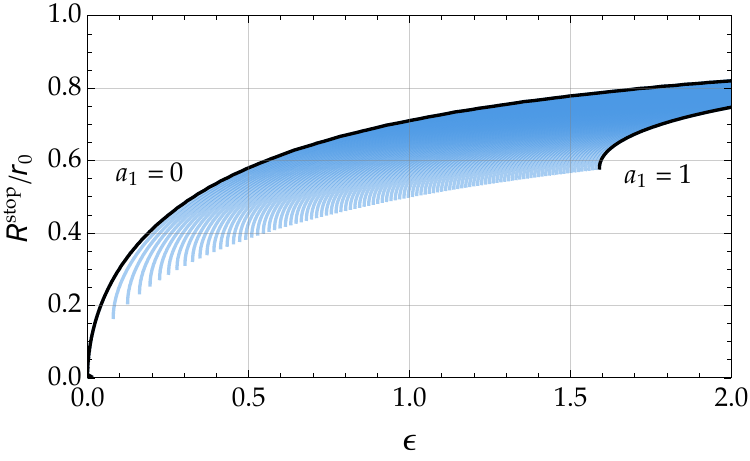}
    \caption{Dependence of $R^{\rm stop}$ on the spacetime parameters of the parametrized spacetime.}
    \label{fig-rstop}
\end{figure}

In this section, we consider the motion of the particle along the radial coordinate only. In radial geodesics, the particle angular velocities of the particle vanish ($u^\theta=0=u^\phi$). In eq. (\ref{ut}) we are already given the explicit form of the time component of the four-velocity. The only component of the four-velocity we need to determine is the radial component. To find it, we use normalization condition $u^\mu u_\mu=-1$ for the massive particle and obtain the following expression:
\begin{eqnarray}\label{ur2}
    B^2(u^r)^2=E^2-N^2(r)\ .
\end{eqnarray}
If one considers the radially falling particle from the rest position at $r=b$, from (\ref{ur2}) one can find that the energy of the particle equals
\begin{eqnarray}\label{at-rest-energy}
    E=N(b)\ .
\end{eqnarray}
It has been shown in \cite{2006gere:Hobson,Crispino:2016pnv} that in certain spacetimes the freely falling particle stops at some point and bounces back. This point of the spacetime (denoted as $r=R^{\rm stop}$) is found from the conservation of energy given by eq. (\ref{ur2}). Therefore, we can claim that the radius $R^{\rm stop}$ does not depend on the spacetime parameters $b_i$. Since one cannot find the explicit form of $R^{\rm stop}$ for the parametrized spacetime with metric functions (\ref{metric-functions}), we express the dependence of this radius on the spacetime parameters in Fig. \ref{fig-rstop}. Fig. \ref{fig-rstop} shows that with increasing the value of the spacetime parameter $a_1$, the radius $R^{\rm stop}$ decreases.
If the spacetime parameter $a_1$ is neglected, one can find analytical expression of $R^{\rm stop}$ as
\begin{eqnarray}\label{r-stop}
    R^{\rm stop}=\frac{2br_0 \sqrt{\epsilon }}{\sqrt{4 b^2 (\epsilon +1)-3 r_0^2 \epsilon }-r_0\sqrt{\epsilon}}\ .
\end{eqnarray}
From (\ref{r-stop}) and Fig. \ref{fig-rstop} one can see that in the case of the Schwarzschild black hole ($\epsilon=0$, $a_1=0$), the particle does not bounce back and travels towards the curvature singularity of the spacetime as $R^{\rm stop}=0$ \cite{2006gere:Hobson}. If the initial position of the particle is at spatial infinity ($r\rightarrow\infty$), the radius $R^{\rm stop}$ is located at
\begin{eqnarray}
  R^{\rm stop}=r_0\sqrt{\frac{\epsilon}{1+\epsilon}}\ ,
\end{eqnarray}
which is located inside the event horizon of the black hole ($R^{\rm stop}<r_0$). If the initial position of the particle is very far relative to the horizon of the black hole, the radius $R^{\rm stop}$ is located at 
\begin{eqnarray}
  R^{\rm stop}=r_0\sqrt{\frac{\epsilon}{1+\epsilon}}+\left(\frac{\epsilon}{1+\epsilon}\right)\frac{r_0^2}{2b}+O\left(b^{-2}\right)\ .  
\end{eqnarray}
In this case also the radius $R^{\rm stop}$ is located inside the event horizon of the black hole. Since $R^{\rm stop}$ is always located inside the event horizon of the black hole and we are interested in the tidal forces outside the event horizon, we will not explore the bounce radius in this paper.

Let us write the motion of the particle in terms of Newton's second law in which the acceleration of the particle equals the net force per mass. It is well known that the radial acceleration of the particle is defined by the second derivative of radial coordinate (or the first derivative of radial velocity) with respect to the proper time. Thus, the "Newtonian radial acceleration" that the parametrized black hole exerts on the radially freely falling particle is given by
\begin{eqnarray}\label{accel}
    \dot{u}^r=
    -\frac{r_0}{2r^2}\left[1+\epsilon\left(1-\frac{3r_0^2}{r^2}\right)-\frac{r_0^2}{r^2}\left(3-\frac{4r_0}{r}\right)a_1\right].
\end{eqnarray}
\begin{figure*}[th]
\centering
\includegraphics[width=0.47\textwidth]{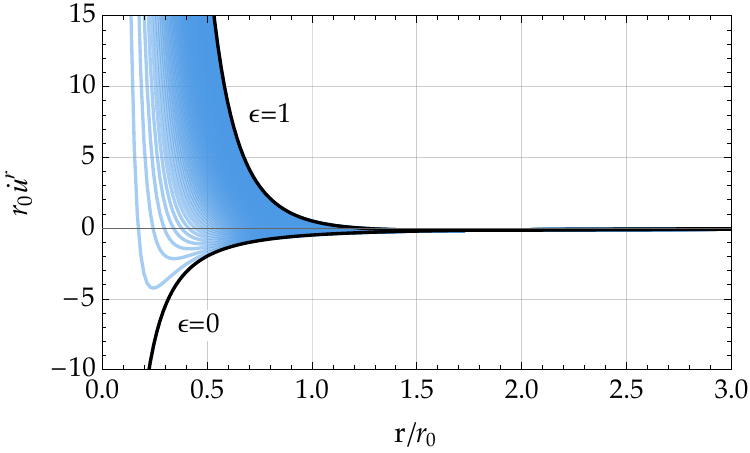}
\includegraphics[width=0.47\textwidth]{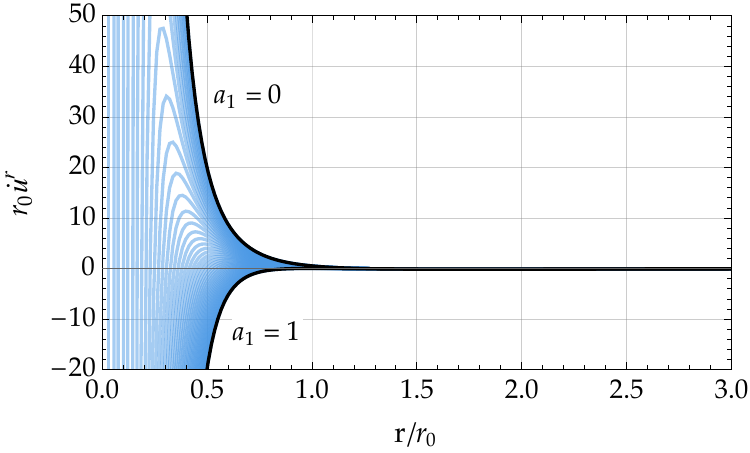}
\caption{The "Newtonian radial acceleration" of the particle in parametrized spacetime. Where we set $a_1=0=b_1$ and $\epsilon\in[0,1]$ with 0.01 steps (on the left panel) and $\epsilon=1$ and $a_1\in[0,1]$ with 0.01 steps (on the right panel).}
\label{fig-accel}
\end{figure*}
Indeed, in the case of the Schwarzschild black hole, the Newtonian radial acceleration reduces to the free fall acceleration in Newtonian gravity \cite{2006gere:Hobson,Mahajan:NCBS}. However, in the case of the parametrized black hole due to the effect of  the additional parameter $\epsilon$, its behaviour changes significantly, see Fig. \ref{fig-accel} for the details.

\section{Geodesic deviation in parametrized spacetime}\label{sec-geod-dev}

Now let us focus on the tidal force acting on the radially freely falling particle to the black hole in the parametrized spacetime. It is well known that the distance between two freely falling particles is described by the geodesic deviation vector $\eta^\mu$ via the equation
\begin{eqnarray}\label{geo-dev}
    \frac{D^2\eta^\mu}{D\tau^2}=R^\mu_{\ \sigma\nu\rho}v^\sigma v^\nu \eta^\rho\ ,
\end{eqnarray}
where $\tau$ is a proper time, $v^\mu$ is the unit vector tangent to the geodesics. The geodesic deviation refers to the phenomenon where initially parallel geodesics (the paths followed by freely falling particles) start to converge or diverge due to the curvature of spacetime. For our further calculations, we turn to the free-fall reference frame via the following orthonormal tetrads:
\begin{eqnarray}\label{tetrad}
    &&\hat{e}_{\hat 0}^{\ \mu}=\frac{E}{N^2(r)}\delta^{(\mu)}_{(0)}-\frac{\sqrt{E^2-N^2(r)}}{B(r)}\delta^{(\mu)}_{(1)}\ ,\nonumber\\
    &&\hat{e}_{\hat 1}^{\ \mu}=-\frac{\sqrt{E^2-N^2(r)}}{B^2(r)}\delta^{(\mu)}_{(0)}+\frac{E}{B(r)}\delta^{(\mu)}_{(1)}\ ,\\
    &&\hat{e}_{\hat 2}^{\ \mu}=\frac{1}{r}\delta^{(\mu)}_{(2)}\ ,\qquad \hat{e}_{\hat 3}^{\ \mu}=\frac{1}{r\sin\theta}\delta^{(\mu)}_{(3)}\ ,\nonumber
\end{eqnarray}
where indices $(0,1,2,3)=(t,r,\theta,\phi)$. These tetrads satisfy the orthonormality condition as
\begin{eqnarray}
    \hat{e}_{\hat \mu}^{\ \alpha}\hat{e}_{\hat{\nu} \alpha}=\eta_{\hat\mu\hat\nu}\ ,
\end{eqnarray}
where $\eta_{\hat\mu\hat\nu}$ is the metric tensor of Minkowski metric. Note that $\hat{e}_{\hat 0}^\mu=u^\mu$. Let us write the geodesic deviation vector in terms of the free-fall reference frame tetrads as
\begin{eqnarray}\label{eta-hat}
    \eta^\mu=\hat{e}_{\hat\nu}^{\ \mu} \eta^{\hat\nu}\ ,
\end{eqnarray}
where we set $\eta^{\hat\nu}=(0,\eta^{\hat 1},\eta^{\hat 2},\eta^{\hat 3})$. In terms of the free-fall reference frame, the radial geodesic deviation equation (\ref{geo-dev}) takes the following form:
\begin{eqnarray}
    \frac{D^2\eta^{\hat 1}}{D\tau^2}=-\frac{1}{2B}\frac{d}{dr}\left(\frac{1}{B}\frac{dN^2}{dr}\right)\eta^{\hat 1}\ ,\label{geo-dev-rad}
\end{eqnarray}
while the angular geodesic deviation equation is given by
\begin{eqnarray}
        \frac{D^2\eta^{\hat i}}{D\tau^2}=-\frac{1}{rB^2}\left[\frac{1}{2}\frac{dN^2}{dr}+\frac{E^2-N^2}{B}\frac{dB}{dr}\right]\eta^{\hat i}\ ,\label{geo-dev-ang}
\end{eqnarray}
where $i=2,3$. For the exterior spherically symmetric spacetimes with $g_{tt}g_{rr}=-1$ (or $B(r)=1$), the geodesic deviation equations (\ref{geo-dev-rad}) and (\ref{geo-dev-ang}) reduce to the well-known forms \cite{Lima:2020wcb,Sharif:2018gzj}
\begin{eqnarray}
    &&\frac{D^2\eta^{\hat 1}}{D\tau^2}=-\frac{1}{2}\frac{d^2N^2}{dr^2}\eta^{\hat 1}\ ,\\
    &&\frac{D^2\eta^{\hat i}}{D\tau^2}=-\frac{1}{2r}\frac{dN^2}{dr}\eta^{\hat i}\ .
\end{eqnarray}
In the next sections, we analyze radial and angular tidal forces of the radially freely falling particle in detail.

\section{Radial tidal force}\label{sec-rtf}

\begin{figure*}[th]
\centering
\includegraphics[width=0.47\textwidth]{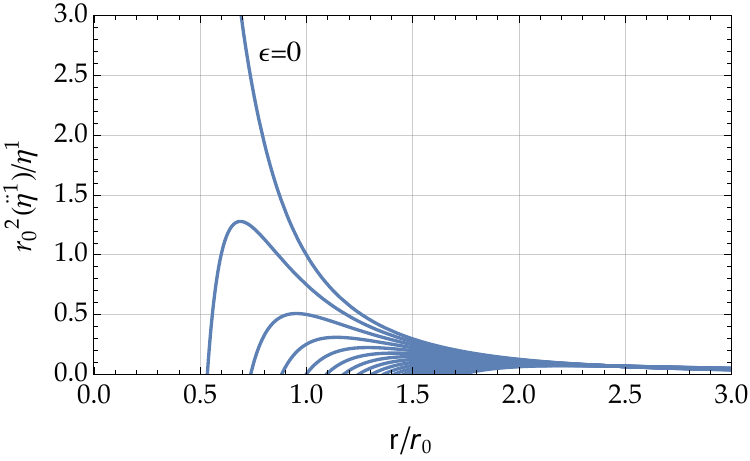}
\includegraphics[width=0.47\textwidth]{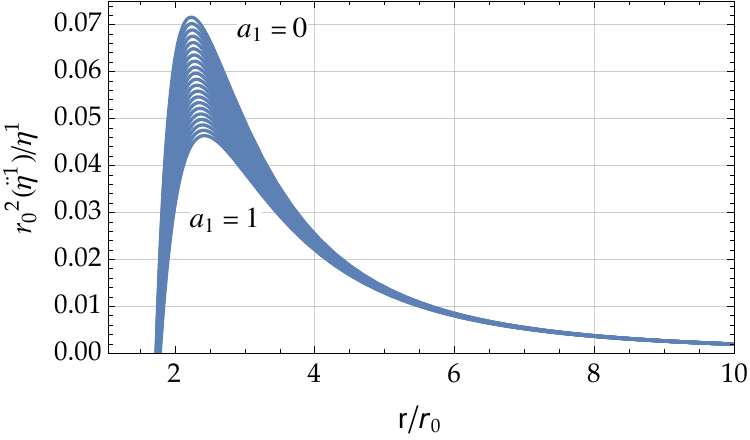}
\caption{The radial profile of the radial force for different values of the spacetime parameters. Where we set $a_1=0=b_1$ and $\epsilon\in[0,1]$ with 0.05 steps (on the left panel) and $\epsilon=1$ and $a_1\in[0,1]$ with 0.05 steps (on the right panel).}
\label{fig-Frtf-radial}
\end{figure*}

In this section, we explore the radial tidal force on the neutral test particle radially falling to a parametrized black hole with spacetime (\ref{line-element}). Let us first consider the radial profile of the radial tidal force (\ref{geo-dev-rad}) which is given in Fig. \ref{fig-Frtf-radial}.  
As it can be seen in Fig. \ref{fig-Frtf-radial} that in the Schwarzschild spacetime, the radial tidal force always increases towards center of the spacetime. However, in the case of the parametrized spacetime, close to the center of the spacetime the radial tidal force is negative and with increasing the radial coordinate, it increases to the positive maximum value and beyond decreases to zero and finally vanishes. An increase in the value of $\epsilon$ increases the region where the radial tidal force becomes negative as well as decreases the maximum value of the radial tidal force. The parameter $a_1$, on the other hand, does not affect significantly the radius inside of which the tidal force is negative, while it slightly decreases the maximum value of the radial tidal force. 

\begin{figure*}[th]
\centering
\includegraphics[width=0.47\textwidth]{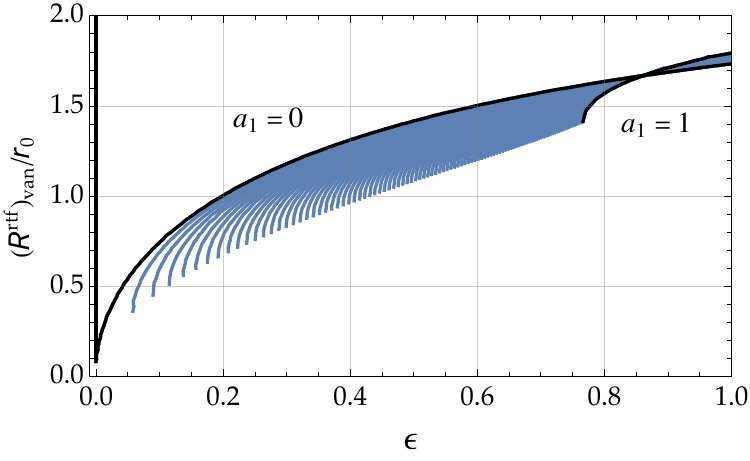}
\includegraphics[width=0.47\textwidth]{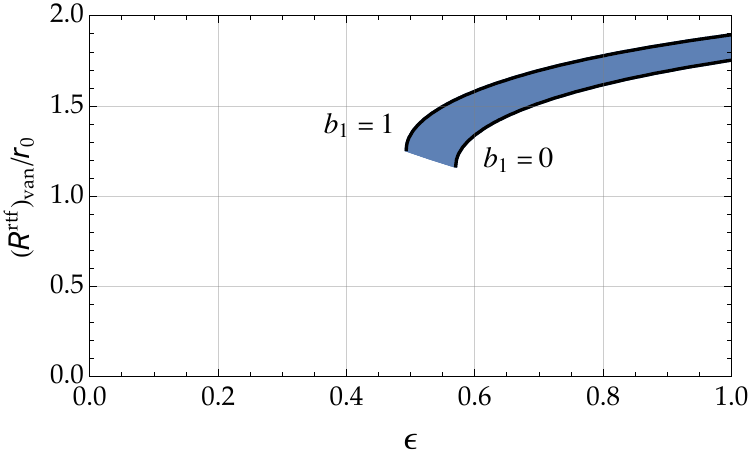}
\caption{The radius where the radial force vanishes for different values of the spacetime parameters. Where we set $b_1=0$ and $a_1\in[0,1]$ with 0.01 steps (on the left panel) and $a_1=0.5$ and $b_1\in[0,1]$ with 0.01 steps (on the right panel).}
\label{fig-Frtf-vanish}
\end{figure*}

As we have seen in Fig. \ref{fig-Frtf-radial}, the radial tidal force is negative close to a center of the parametrized spacetime and it is positive elsewhere. Between these positive and negative values of the radial tidal force, at some point in the radial coordinate, it becomes zero. From (\ref{geo-dev-rad}) one can determine the radius $r=R^{\rm rtf}_{\rm van}$ where the radial tidal force vanishes by solving the following polynomial equation:
\begin{eqnarray}\label{radial-rtf-eq}
    &&(\epsilon+1)\left(R^{\rm rtf}_{\rm van}\right)^5-6r_0^2(\epsilon+a_1) \left(R^{\rm rtf}_{\rm van}\right)^3 \\ 
    &&+10a_1r_0^3\left(R^{\rm rtf}_{\rm van}\right)^2-3b_1r_0^4(\epsilon+a_1)R^{\rm rtf}_{\rm van}+6a_1b_1r_0^5=0\ .\nonumber
\end{eqnarray}
The polynomial equation (\ref{radial-rtf-eq}) cannot be solved analytically. Therefore, in order to clarify the effects of the spacetime parameters on the radius $R^{\rm rtf}_{\rm van}$ we solve it numerically and present results in Fig. \ref{fig-Frtf-vanish}. Examining Fig. \ref{fig-Frtf-vanish} might lead one to conclude that the radial tidal force vanishes at the center of the Schwarzschild spacetime. However, this interpretation contradicts the result depicted in Fig. \ref{fig-Frtf-radial}, which illustrates that in the Schwarzschild black hole scenario, the tidal force diverges as the particle approaches the curvature singularity at $r=0$. To clarify this discrepancy, let us analyze the matter further. In Fig. \ref{fig-Frtf-vanish}, the black curve corresponding to $a_1=0$ denotes the points at which the radial tidal force vanishes. The expression for this force is given by:
\begin{eqnarray}\label{force-epsilon}
    \frac{r_0^2\ddot{\eta}^{\hat 1}}{\eta^{\hat 1}}=\frac{(1+\epsilon)r_0}{r^3}-\frac{6\epsilon r_0^3}{r^5}\ .
\end{eqnarray}
From (\ref{force-epsilon}) it is evident that the radial tidal force always diverges as the particle approaches the spacetime center, even when $\epsilon=0$. Nonetheless, with any positive $\epsilon$ value, the second fraction on the right-hand side of (\ref{force-epsilon}) prevails for small $r$ values, leading to the radial tidal force reaching negative infinity at $r=0$. Thus, if the spacetime deviates from the Schwarzschild one on account of the parameters $\epsilon$ and $a_1$, the radius $R^{\rm rtf}_{\rm van}$ becomes non-zero and with increasing the value of the parameter $\epsilon$, the radius $R^{\rm rtf}_{\rm van}$ increases. On the contrary, the parameter $a_1$ decreases $R^{\rm rtf}_{\rm van}$. On the other hand, the parameter $b_1$ does not significantly affect to the radius $R^{\rm rtf}_{\rm van}$, however its influence is positive, as with increasing the value of $b_1$, the radius where the radial tidal force vanishes increases.
As we mentioned above, once a deviation from general relativity is considered, the radius $R^{\rm rtf}_{\rm van}$ becomes non-zero and depending on the values of the spacetime, it can be inside or outside the event horizon. When the spacetime parameters satisfy the following condition, the radial tidal force vanishes at the event horizon of the spacetime:
\begin{eqnarray}
    \epsilon_{\rm cr}=a_1+\frac{1-a_1}{5+3b_1}\ .
\end{eqnarray}
Accordingly, the radius $R^{\rm rtf}_{\rm van}$ relative to the event horizon can be classified as follows:
\begin{itemize}
    \item[(i)] For $\epsilon<\epsilon_{\rm cr}$, the radius $R^{\rm rtf}_{\rm van}$ is located inside the event horizon, $R^{\rm rtf}_{\rm van}<r_0$;
    \item[(ii)] For $\epsilon=\epsilon_{\rm cr}$, the radius $R^{\rm rtf}_{\rm van}$ is located at the event horizon, $R^{\rm rtf}_{\rm van}=r_0$;
    \item[(iii)] For $\epsilon>\epsilon_{\rm cr}$, the radius $R^{\rm rtf}_{\rm van}$ is located outside the event horizon, $R^{\rm rtf}_{\rm van}>r_0$.
\end{itemize}
If the parameter $a_1$ is negligibly small, then the radius $R^{\rm rtf}_{\rm van}$ becomes
\begin{eqnarray}\label{radial-rtf}
    R^{\rm rtf}_{\rm van}=r_0\left[\frac{3\epsilon}{\epsilon+1}+\frac{\sqrt{3\epsilon} \sqrt{b_1(\epsilon+1)+3\epsilon}}{\epsilon +1}\right]^{1/2}.
\end{eqnarray}
Now let us turn our focus on the maximum value of the radial tidal force exerted by the black hole to the radially falling free particle. Prior to determining the maximum value of the radial tidal force, let us find the radius $r=R^{\rm rtf}_{\rm max}$ where the tidal force reaches the maximum value. To find this radius, one must solve the below given polynomial equation
\begin{eqnarray}\label{r-rtf-max}
    &&(\epsilon+1)\left(R^{\rm rtf}_{\rm max}\right)^5-10r_0^2(a_1+\epsilon) \left(R^{\rm rtf}_{\rm max}\right)^3\\
    && +20a_1r_0^3\left(R^{\rm rtf}_{\rm max}\right)^2-7b_1 r_0^4(a_1+\epsilon)R^{\rm rtf}_{\rm max}+16 a_1b_1r_0^5=0\ .\nonumber
\end{eqnarray}
Unfortunately, the fifth order polynomial equation (\ref{r-rtf-max}) cannot be solved analytically for the radius $R^{\rm rtf}_{\rm max}$.
\begin{figure*}[th]
\centering
\includegraphics[width=0.47\textwidth]{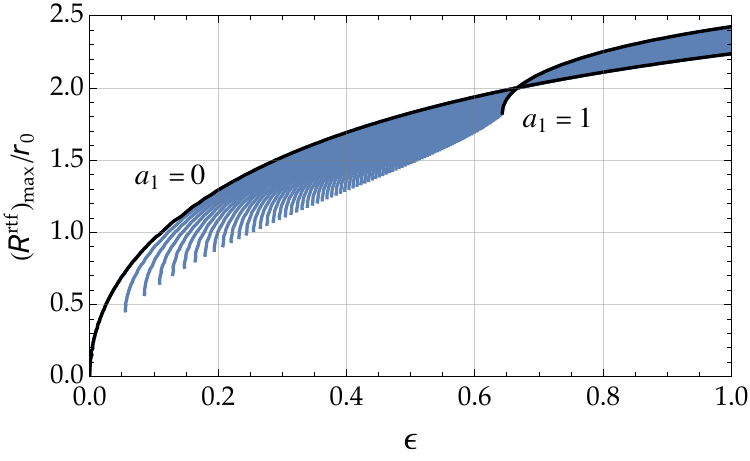}
\includegraphics[width=0.47\textwidth]{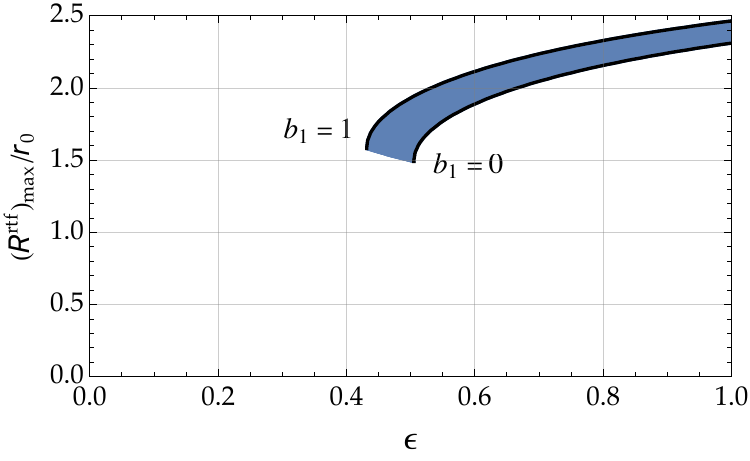}
\caption{The radius where the radial radial force becomes maximum as a function of the parameters of the parametrized spacetime. Where we set $b_1=0$ and $a_1\in[0,1]$ with 0.01 steps (on the left panel) and $a_1=0.5$ and $b_1\in[0,1]$ with 0.01 steps (on the right panel).}
\label{fig-Rrtf-max}
\end{figure*}
By solving eq. (\ref{r-rtf-max}) numerically, one can easily determine the effect of the parameters of the spacetime on the radius. In Fig. \ref{fig-Rrtf-max} we present some special cases of the effects of parameters $\epsilon$, $a_1$ and $b_1$. Moreover, in order to have more precise knowledge of the relation of the spacetime parameters to $R^{\rm rtf}_{\rm max}$, we found the approximate analytical solution of eq. (\ref{r-rtf-max}) by assuming the higher order parameters $a_1$ and $b_1$ are very small as
\begin{eqnarray}\label{r-rthf-max0}
    R^{\rm rtf}_{\rm max}=&&r_0\left[\sqrt{\frac{10\epsilon}{1+\epsilon}}-\left(\frac{1}{\epsilon}-\sqrt{\frac{5}{2\epsilon(1+\epsilon)}}\right)a_1\right.\nonumber\\
    && \left.+\frac{7}{20}\sqrt{\frac{1+\epsilon}{10\epsilon}}b_1\right]+O(a_1^2, b_1^2)\ .
\end{eqnarray}
From Fig. \ref{fig-Rrtf-max} and eq. (\ref{r-rthf-max0}) one can see that with increasing the value of $\epsilon$, the radius $R^{\rm rtf}_{\rm max}$ always increases but on account of only parameter $\epsilon$, the radius $R^{\rm rtf}_{\rm max}$ can be located outside the event horizon only if $\epsilon>1/9$. Moreover, the parameter $a_1$ decreases the radius $R^{\rm rtf}_{\rm max}$ while, the parameter $b_1$ increases this radius however its contribution is not much significant.

Now let us turn our attention on the maximum radial tidal force acting on the radially freely falling particle to the parametrized black hole. To calculate the maximum value of the radial tidal force (\ref{geo-dev-rad}), we use equation (\ref{r-rtf-max}) and find the following function:
\begin{eqnarray}\label{t-force-rad-max}
    \left(\frac{\ddot{\eta}^{\hat 1}}{\eta^{\hat 1}}\right)_{{\rm max}}=\frac{2r_0^3\left[2(\epsilon+a_1)R^{\rm rtf}_{\rm max}-5a_1r_0\right]}{(R^{\rm rtf}_{\rm max})^2\left[(R^{\rm rtf}_{\rm max})^2+b_1r_0^2\right]^2}\ .
\end{eqnarray}
\begin{figure*}[th]
\centering
\includegraphics[width=0.47\textwidth]{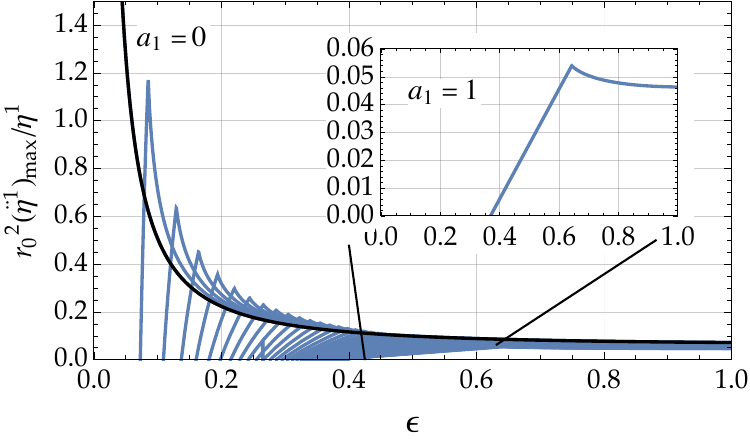}
\includegraphics[width=0.47\textwidth]{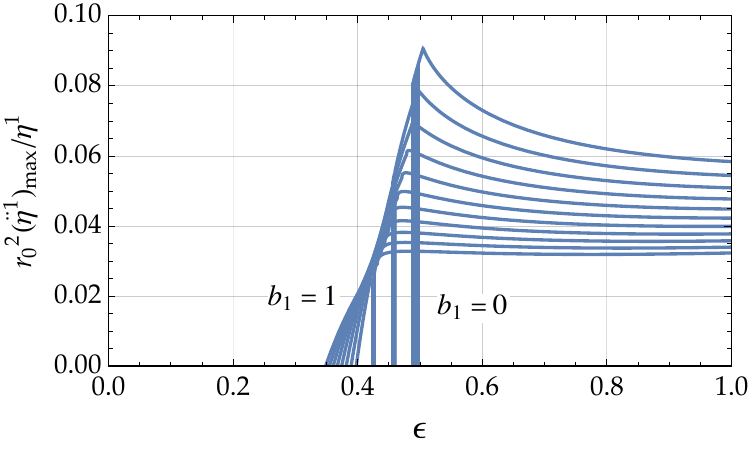}
\caption{The maximum radial force as a function of the parameters of the parametrized spacetime. Where we set $b_1=0$ and $a_1\in[0,1]$ with 0.05 steps (on the left panel) and $a_1=0.5$ and $b_1\in[0,1]$ with 0.1 steps (on the right panel).}\label{fig-Frtf-max}
\end{figure*}
To calculate exact maximum value of the radial tidal force given by (\ref{t-force-rad-max}), one must find the value of the radius $R^{\rm rtf}_{\rm max}$ in eq. (\ref{r-rtf-max}). By solving eq. (\ref{r-rtf-max}) numerically and inserting that values to the maximum value of the radial tidal force (\ref{t-force-rad-max}), we obtain the numerical results of which some cases presented in Fig. \ref{fig-Frtf-max}. From Fig. \ref{fig-Frtf-max} one can see that in the Schwarzschild black hole spacetime, the maximum value of the radial tidal force diverges at $r=0$. However, in the parametrized spacetime the tidal force never becomes infinity. If we consider the effect of a single parameter $\epsilon$, it decreases the maximum value of the radial tidal force. 

However, an approximate maximum value of the radial tidal force for the small $a_1$ and $b_1$ parameters can easily be calculated via (\ref{r-rthf-max0}) as
\begin{eqnarray}\label{rtf-max-approx}
   \left(\frac{\ddot{\eta}^{\hat 1}}{\eta^{\hat 1}}\right)_{{\rm max}}&&=\frac{1}{25}\left\{\frac{(\epsilon +1)^{5/2}}{\sqrt{10} \epsilon ^{3/2}}-\frac{3 b_1(\epsilon +1)^{7/2}}{8\sqrt{10}\epsilon^{5/2}}\right.\\
   &&\left.+ a_1\left[\frac{(\epsilon +1)^3}{4\epsilon^3}-\frac{3 (\epsilon +1)^{5/2}}{2 \sqrt{10}\epsilon ^{5/2}}\right] \right\}\frac{1}{r_0^2}+O(a_1^2, b_1^2).\nonumber
\end{eqnarray}
From (\ref{rtf-max-approx}) one can confirm that the results of Fig. \ref{fig-Frtf-max} that the parameter $a_1$ increases the maximum value of the radial tidal force while, the parameter $b_1$ decreases it.

\begin{figure*}[th]
\centering
\includegraphics[width=0.47\textwidth]{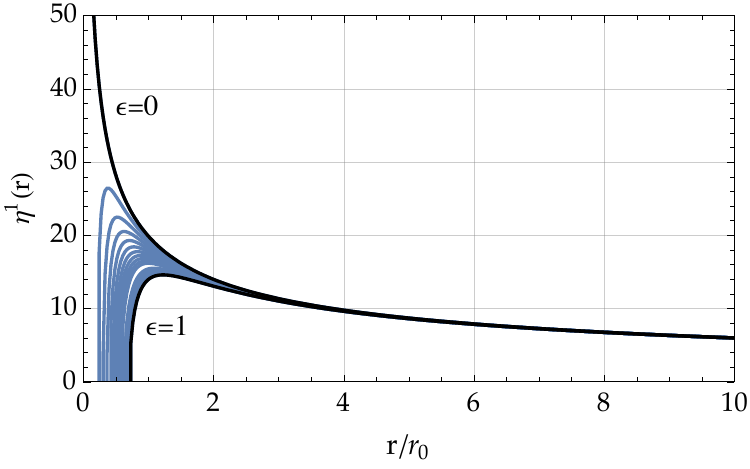}
\includegraphics[width=0.47\textwidth]{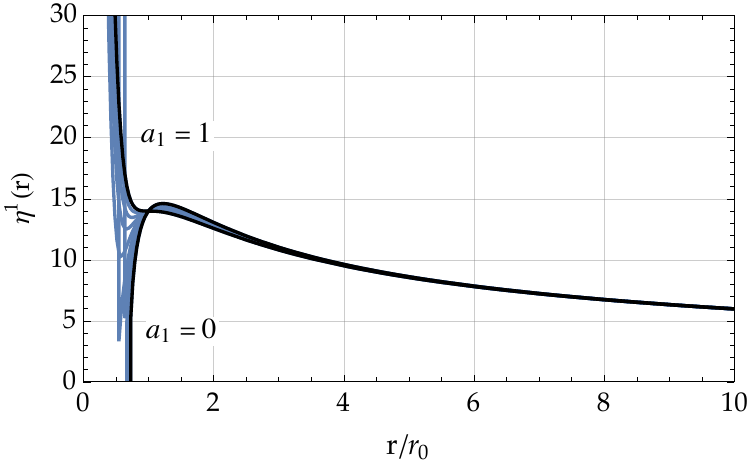}
\caption{The radial profile of the radial component of the geodesic deviation vector $\eta^{\hat 1}$ in the parametrized spacetime. Where we adopted IC1, $b=100r_0$ and set $a_1=0=b_1$, $\epsilon\in[0,1]$ with 0.1 steps (on the left panel) and $\epsilon=1$, $a_1\in[0,1]$ with 0.1 steps (on the right panel).}
\label{fig-eta-r-ic1}
\end{figure*}

\begin{figure*}[th]
\centering
\includegraphics[width=0.47\textwidth]{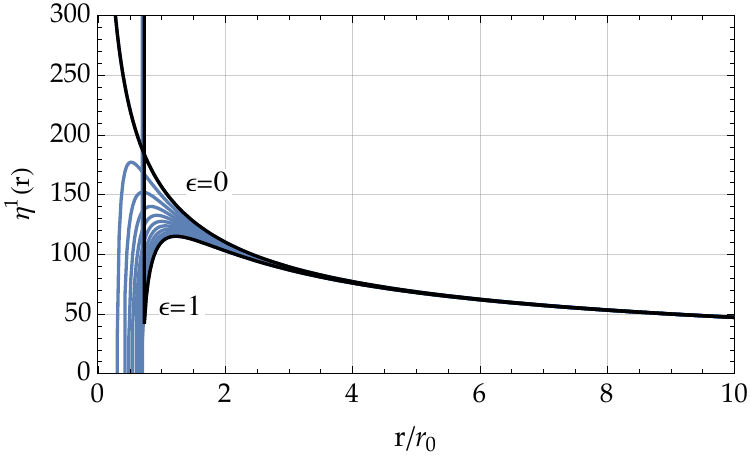}
\includegraphics[width=0.47\textwidth]{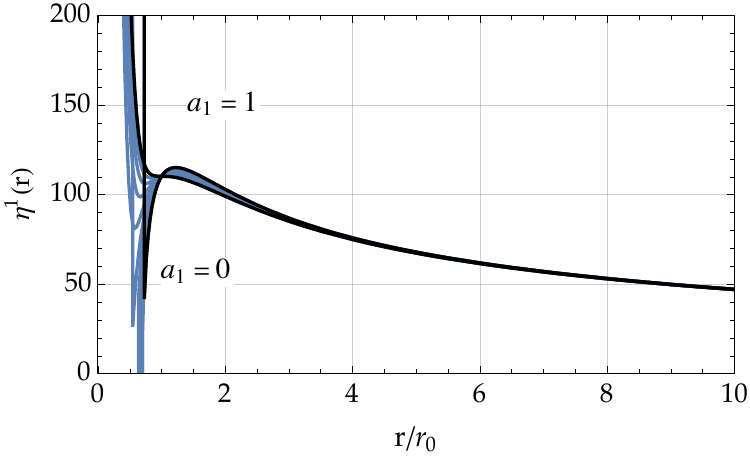}
\caption{The same as Fig. \ref{fig-eta-r-ic1} but for the IC2.}
\label{fig-eta-r-ic2}
\end{figure*}

Now we solve the equation of geodesic deviation (\ref{geo-dev-rad}) for the radial geodesic deviation vector $\eta^{\hat 1}$ in the parametrized spacetime (\ref{line-element}) with metric functions (\ref{metric-functions}). First, we write the differential equation with respect to the radial coordinate through the following relation:
\begin{eqnarray}\label{tau-to-r}
    \frac{d^2\eta^{\hat 1}}{d\tau^2}=(u^r)^2\frac{d^2\eta^{\hat 1}}{dr^2}+\frac{1}{2}\frac{d(u^r)^2}{dr}\frac{d\eta^{\hat 1}}{dr}\ .
\end{eqnarray}
By using expression (\ref{tau-to-r}) and (\ref{ur2}), we rewrite the differential equation for the radial geodesic deviation (\ref{geo-dev-rad}) in the following form:
\begin{eqnarray}\label{geo-dev-rad-dif}
    &&\frac{d^2\eta^{\hat 1}}{dr^2}+\left(\frac{NN'}{N^2-E^2}-\frac{B'}{B}\right)\frac{d\eta^{\hat 1}}{dr}\nonumber\\
    &&+\frac{N \left(B N''-B' N'\right)+B(N')^2}{B\left(E^2-N^2\right)}\eta^{\hat 1}=0\ .
\end{eqnarray}
To solve ordinary differential equation for the radial geodesic deviation (\ref{geo-dev-rad-dif}), we adopt the two types of initial condition for the particle falling from $r=b>r_0$ to the parametrized black hole \cite{Crispino:2016pnv,Hong:2020bdb,LimaJunior:2022gko,LimaJunior:2020fhs}. The first initial conditions (IC1) given by
\begin{eqnarray}\label{IC1}
\eta^{\hat 1}(b)=1\ , \qquad \frac{d\eta^{\hat 1}}{d\tau}|_{r=b}=0\ ,
\end{eqnarray}
represents the particle with no internal motion at $r=b$. The second initial conditions (IC2) represents dust "exploding" from a point at $r=b$ on the symmetry axis via the condition
\begin{eqnarray}\label{IC2}
\eta^{\hat 1}(b)=0\ , \qquad \frac{d\eta^{\hat 1}}{d\tau}|_{r=b}=1\ .
\end{eqnarray}
In the next step we numerically solve the ordinary differential equation (\ref{geo-dev-rad-dif}) in the parametrized spacetime (\ref{line-element}) with metric functions (\ref{metric-functions}) for the initial conditions IC1 and IC2 and present radial dependencies of the radial geodesic deviation vector in Figs. \ref{fig-eta-r-ic1} and \ref{fig-eta-r-ic2}. 

In Fig. \ref{fig-eta-r-ic1} we adopt IC1 and set $b=100r_0$, $a_1=0=b_1$ plot the $\eta^{\hat 1}$ component of the geodesic deviation vectors for various choices of the parameter $\epsilon$ of the parametrized black hole spacetime. We show that the behaviour of $\eta^{\hat 1}$ is essentially the same for different values of $\epsilon$ at large distances. Besides that, it can be seen that the geodesic deviation vector $\eta^{\hat 1}$ in the Schwarzschild spacetime tends to infinity as the particle approaches the center of the spacetime where the curvature singularity is located. However, the behaviour of $\eta^{\hat 1}$ changes with increasing the value of $\epsilon$, as it initially increases till its maximum value (inside the black hole horizon) and it decreases to a very small value at $R^{\rm stop}$ near the spacetime singularity. With increasing the value of $\epsilon$, the maximum value of $\eta^{\hat 1}$ decreases and the radial position where it becomes maximum increases towards the black hole horizon, $r=r_0$. In other words, the behaviour of the radial tidal force acting on the particle falling to the black hole is initially stretching and it changes to compressing inside the event horizon of the black hole. In order to explore the effect of the parameter $a_1$, on the right panel of Fig. \ref{fig-eta-r-ic1} we set $\epsilon=1$ and considered the various choices of the parameter $a_1$. In various values of $a_1$ also, the radial geodesic deviation vector behaves the same at large distances but at small distances its effect is significant. In the case of the nonzero values of $\epsilon$ and $a_1$, $\eta^{\hat 1}$ has a local maximum at about the black hole horizon and a local minimum inside the horizon. As it can be seen in Fig. \ref{fig-eta-r-ic1}, the effect of $a_1$ is opposite to $\epsilon$, as with increasing the value of $a_1$, $\eta^{\hat 1}$ does not tend to zero at small distances, instead it reaches a local minimum and tends to positive infinity at small distances. The value of the local minimum of $\eta^{\hat 1}$ increases with increasing the value of $a_1$. To be more precise, a radially infalling particle to the parametrized black hole experiences a radially stretching tidal force throughout its motion until it's inside the horizon. However, once it encounters the horizon, it experiences a radially compressing tidal force. As the value of the parameter $\epsilon$ increases, the particle's radial size becomes almost zero just inside the event horizon. On the other hand, the parameter $a_1$ counteracts the impact of $\epsilon$ by causing the behavior of the radial tidal force to become stretching within the event horizon.

In the case of the second initial conditions (IC2) -- see Fig. \ref{fig-eta-r-ic2}, the qualitative picture of the evolution of the radial geodesic vector with radius does not change significantly that the one with IC1 for the various values of parameters $\epsilon$ and $a_1$, but qualitatively, it changes, as its value becomes greater than in the case of IC1.

\section{Angular tidal force}\label{sec-atf}

\begin{figure*}[th]
\centering
\includegraphics[width=0.47\textwidth]{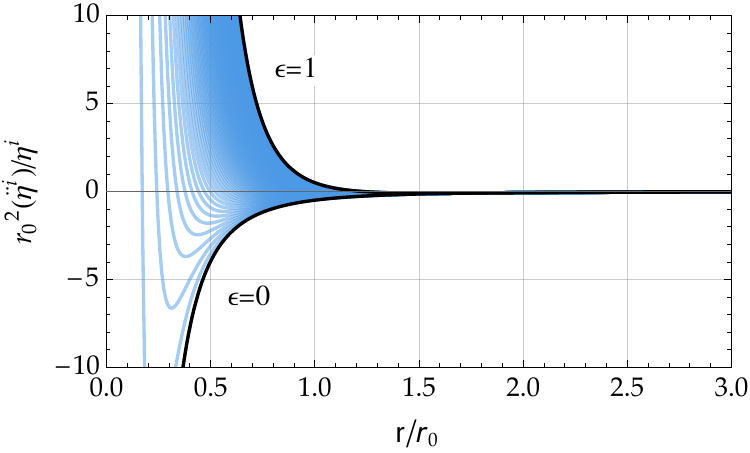}
\includegraphics[width=0.47\textwidth]{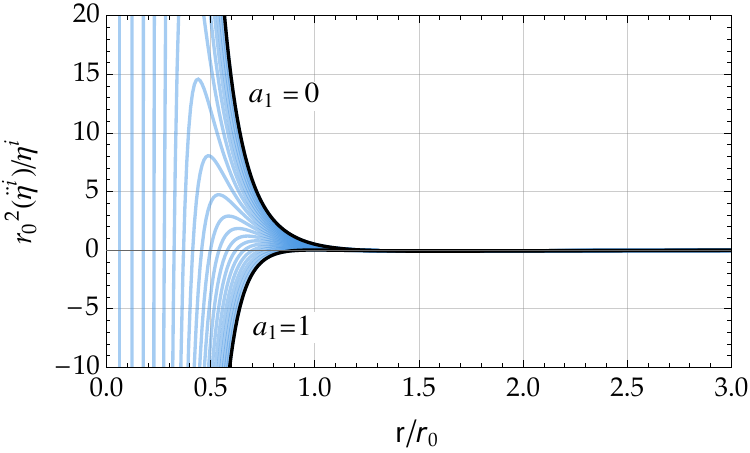}
\caption{The same as Fig. \ref{fig-Frtf-radial} but for the angular tidal force. Where we set $b=100r_0$.}
\label{fig-Fatf-radial}
\end{figure*}

In this section we repeat the calculations presented in the previous section, but for the angular geodesic deviation equation (\ref{geo-dev-ang}). Unlike the radial geodesic equation (\ref{geo-dev-rad}), the evolution of the angular geodesic deviation vector $\eta^{\hat i}$  depends also on the total energy of the particle $E$. Prior to exploring further properties of the angular tidal force, we need to fix the problem associated with the energy of the particle. If one considers that the particle is at rest before radially falling to the black hole, then from the symmetry of the spacetime we can use the relation (\ref{at-rest-energy}) for the energy of the particle. Thus, to study further properties of the angular geodesic equation (or analogously angular tidal force), let us inspect behaviour of the right hand side of the equation (\ref{geo-dev-ang}) with radial coordinates for the various values of the spacetime parameters. In Fig. \ref{fig-Fatf-radial}, we present some patterns of this analysis.
From Fig. \ref{fig-Fatf-radial} one can see that if the particle is positioned at a large distance from the black hole, irrespective of values of the spacetime parameters the angular tidal force exerted by the black hole vanishes as in the case of the radial tidal force. In Schwarzschild spacetime, the angular tidal force becomes negative infinity as one approaches the center of the spacetime. Once the value of $\epsilon$ becomes non-zero, it reaches the minimum finite value inside the horizon of the spacetime and after it diverges to positive infinity towards the center of the spacetime. With increasing the value of $\epsilon$, the minimum value of the angular tidal force decreases and its position tends to the horizon of the spacetime from inside. On the other hand, the parameter $a_1$ plays the opposite role to $\epsilon$. Once we turn on the parameter $a_1$, the radial profile of the angular tidal force takes a barrier-like shape and afterwards, it rapidly decreases and tends to negative infinity as the radial coordinates approach the center.

As we have seen in Fig. \ref{fig-Fatf-radial}, the angular tidal force vanishes in some values of the spacetime parameters. In the next step, we determine the position where the angular tidal force acting on the radially freely falling particle by the parametrized spacetime vanishes. To determine it, we need to solve the following polynomial equation:
\begin{eqnarray}\label{F-atf-vanish}
    &&(\epsilon+1)\left(R^{\rm atf}_{\rm van}\right)^4-4 b_1\left(E^2-1\right)r_0\left(R^{\rm atf}_{\rm van}\right)^3 \nonumber\\
    &&-3(a_1+b_1\epsilon+b_1+\epsilon) r_0^2\left(R^{\rm atf}_{\rm van}\right)^2+4 a_1r_0^3R^{\rm atf}_{\rm van}\nonumber\\
    &&+b_1r_0^4(a_1+\epsilon)=0\ ,
\end{eqnarray}
In the case of the Schwarzschild spacetime, the above equation results that irrespective value of the total energy of the particle, the angular tidal force vanishes at the center of the spacetime, $R^{\rm atf}_{\rm van}=0$ \cite{2006gere:Hobson}, as in the case of the radial tidal force. If we consider the contribution of the spacetime parameter $\epsilon$, then the angular tidal force becomes zero where the radial tidal force vanishes, $R^{\rm atf}_{\rm van}=R^{\rm rtf}_{\rm van}$, as
\begin{eqnarray}
    R^{\rm atf}_{\rm van}=r_0\sqrt{\frac{3\epsilon}{\epsilon+1}}\ .
\end{eqnarray}
To study behaviours of the remaining spacetime parameters $a_1$ and $b_1$, we solve the polynomial equation (\ref{F-atf-vanish}) numerically. 
\begin{figure*}[th]
\centering
\includegraphics[width=0.47\textwidth]{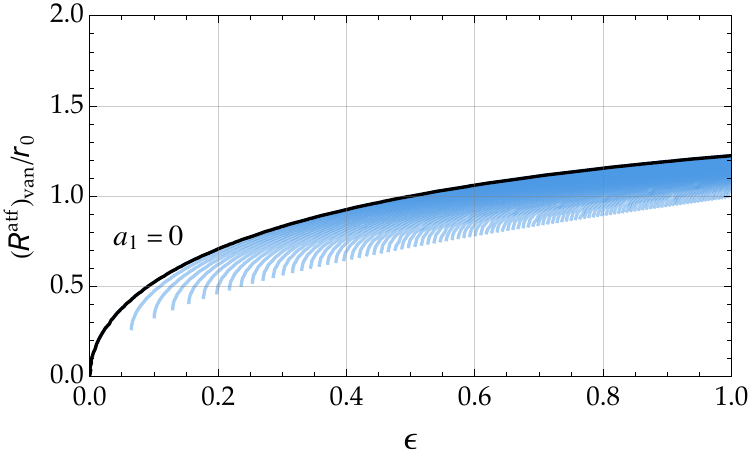}
\includegraphics[width=0.47\textwidth]{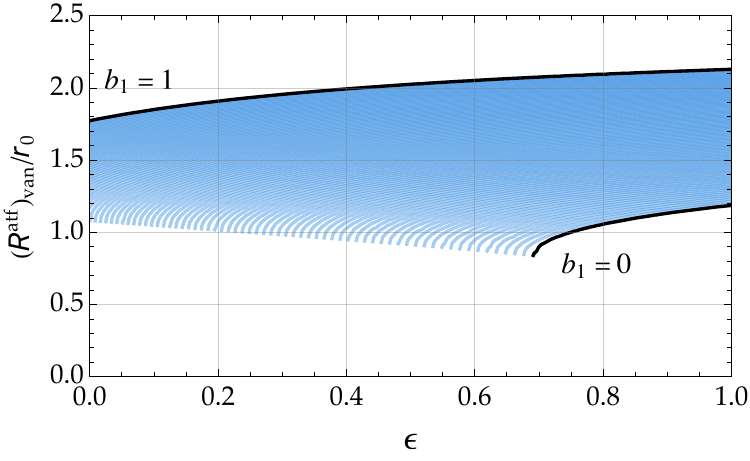}
\caption{The same as Fig. \ref{fig-Frtf-vanish} but for the angular tidal force. Where we set $b=100r_0$.}
\label{fig-Fatf-vanish}
\end{figure*}
In Fig. \ref{fig-Fatf-vanish} we present dependence of $R^{\rm atf}_{\rm van}$ on the spacetime parameters. By comparing Figs. \ref{fig-Fatf-vanish} and \ref{fig-Frtf-vanish}, one can see that there are not many qualitative differences between the dependencies of locations of vanishing the radial and angular tidal forces on the spacetime parameters. However, they differ quantitatively, as the angular tidal force vanishes at closer distances to the black hole than the radial tidal force. Moreover, the effect of the parameter $b_1$ on the angular tidal force is more significant in comparison to the radial case.

Now let us consider the maximum value of the angular tidal force. One can see from Fig. \ref{fig-Fatf-radial} that in the case of the Schwarzschild black hole, the angular tidal force tends to negative infinity as the particle approaches the curvature singularity ($r=0$) and it asymptotically approaches zero as the radius tends to infinity. Once the spacetime deviates from the Schwarzschild spacetime on account of the parameter $\epsilon$, the angular tidal force has an absolute negative minimum, $\left(\frac{\ddot{\eta}^{\hat i}}{\eta^{\hat i}}\right)_{{\rm min}}$, at $r_{\rm min}$ which are in linear approximation of $a_1$ equal to
\begin{eqnarray}
\left(\frac{\ddot{\eta}^{\hat i}}{\eta^{\hat i}}\right)_{{\rm min}}&&=-\frac{(1+\epsilon)^{5/2}}{25 \sqrt{5}r_0^2 \epsilon^{3/2}}\\
&&-\frac{a_1(\epsilon +1)^2 \left(4+4 \epsilon -3 \sqrt{5\epsilon  (\epsilon +1)}\right)}{250 r_0^2 \epsilon ^3}\ ,\nonumber\\
r_{\rm min}&&=\left[\frac{\sqrt{5\epsilon }}{\sqrt{1+\epsilon }}\left(1+\frac{a_1}{2\epsilon}\right)-\frac{4a_1}{5\epsilon}\right]r_0+O(a_1^2),\nonumber\\
\end{eqnarray}
and afterwards, it diverges to positive infinity at the center of the spacetime. For $\epsilon<1/4$, the absolute minimum value of the angular tidal force is located inside the event horizon, $r_{\rm min}<r_0$. For $\epsilon=1/4$ and $\epsilon>1/4$, then $r_{\rm min}=r_0$ and $r_{\rm min}>r_0$, respectively. With increasing the value of $\epsilon$ the minimum value of the angular tidal force decreases and it becomes mostly positive inside the event horizon of the spacetime. However, if we include the effect of the parameter $a_1$, then, as we have mentioned in previous sections, its effect is opposite to $\epsilon$. If the parameter $\epsilon$ makes the angular tidal force diverge to positive infinity at the center, the parameter $a_1$ makes it again diverge to negative infinity at the center of the spacetime. On account of the parameter $a_1$, the angular tidal force has an absolute maximum inside the event horizon of the spacetime. Thus, if the spacetime parameter $a_1$ is not neglected, before tending to negative infinity the angular tidal force has absolute maximum inside the event horizon of the spacetime in the linear approximation of $a_1$ at
\begin{eqnarray}
    r_{\rm max}&&=\frac{8a_1}{5\epsilon}r_0+O(a_1^2)\ ,\\
    \left(\frac{\ddot{\eta}^{\hat i}}{\eta^{\hat i}}\right)_{{\rm max}}&&=\frac{3125 \epsilon ^6}{131072 a_1^5 r_0^2}+\frac{9375 \epsilon ^5}{65536 a_1^4 r_0^2}-\frac{125 (\epsilon +1) \epsilon ^3}{1024 a_1^3 r_0^2}.\nonumber\\
\end{eqnarray}

\begin{figure*}[th]
\centering
\includegraphics[width=0.47\textwidth]{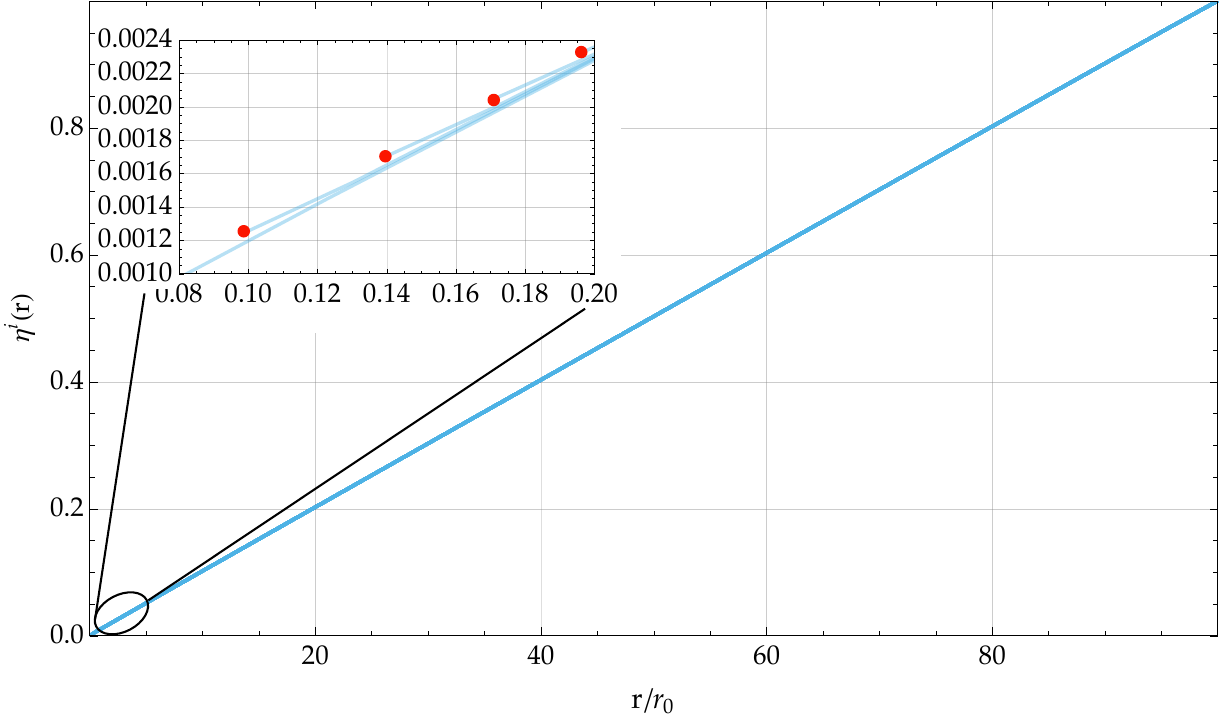}
\includegraphics[width=0.47\textwidth]{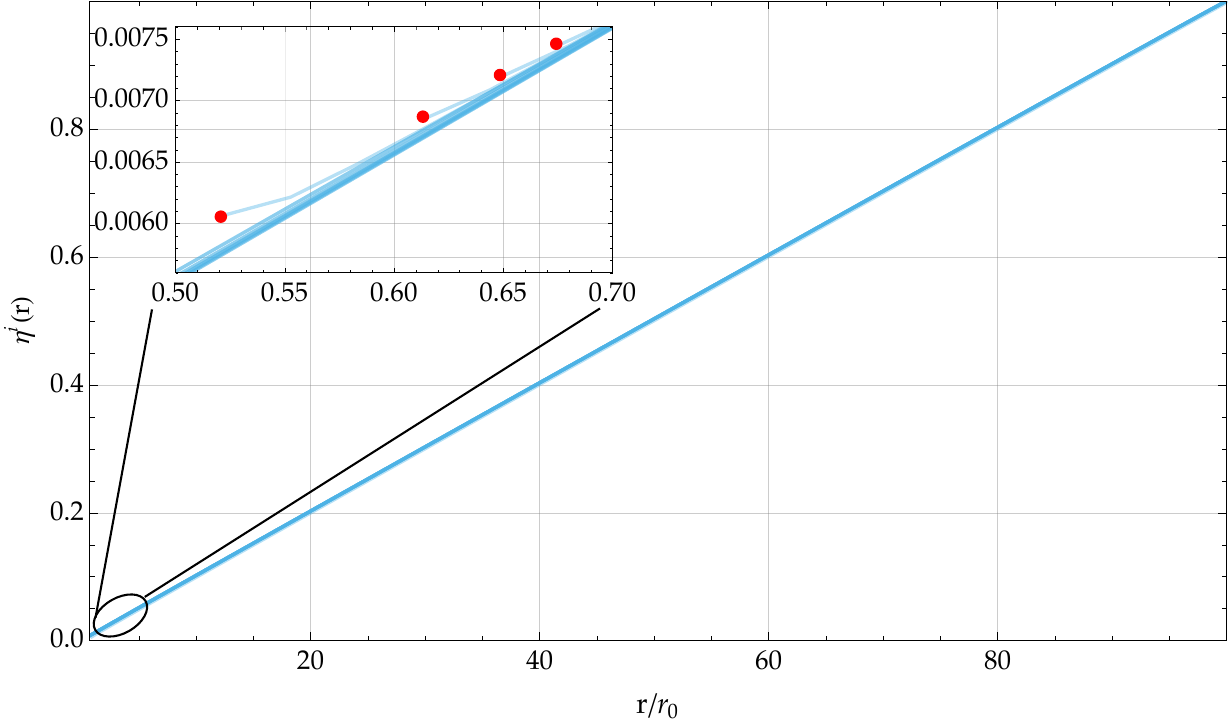}
\caption{The angular profile of the radial component of the geodesic deviation vector $\eta^{\hat i}$ in the parametrized spacetime. Where we adopted IC1, $b=100r_0$ and set $a_1=0=b_1$, $\epsilon\in[0,1]$ with 0.01 steps (on the left panel) and $\epsilon=1$, $b_1=0$, $a_1\in[0,1]$ with 0.01 steps (on the right panel). Where red dots correspond to $R^{\rm stop}$.}
\label{fig-eta-i-ic1}
\end{figure*}

\begin{figure*}[th]
\centering
\includegraphics[width=0.47\textwidth]{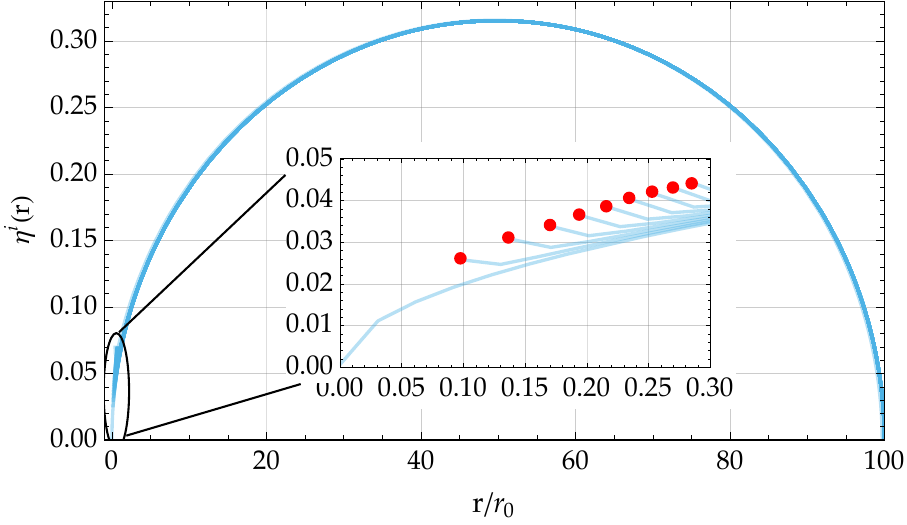}
\includegraphics[width=0.47\textwidth]{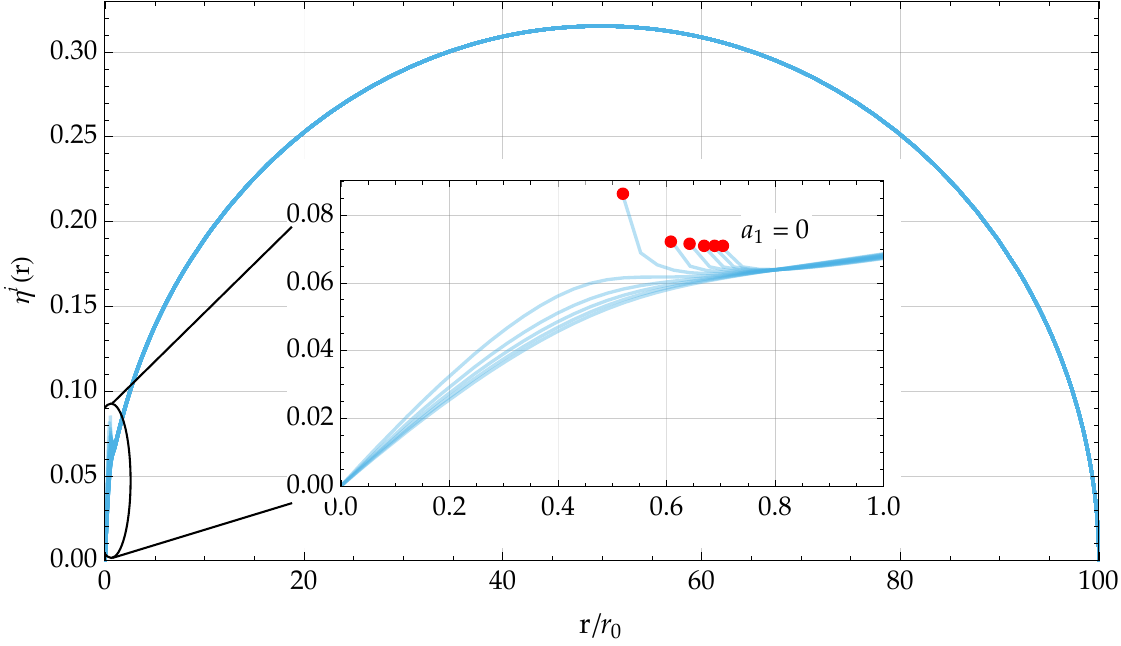}
\caption{The same as Fig. \ref{fig-eta-i-ic1} but for the IC2.}
\label{fig-eta-i-ic2}
\end{figure*}

Let us now focus our attention on the behaviour of the angular component of the deviation vector $\eta^{\hat{i}}$. To investigate the behaviour of the angular deviation vector one needs to solve the second order differential equation (\ref{geo-dev-ang}) with the initial conditions (\ref{IC1}) or (\ref{IC2}). To solve this equation, we first rewrite the differential equation with respect to radial coordinate by using the relation (\ref{tau-to-r}) as
\begin{eqnarray}\label{geo-dev-ang-dif}
    &&\frac{d^2\eta^{\hat i}}{dr^2}+\left(\frac{NN'}{N^2-E^2}-\frac{B'}{B}\right)\frac{d\eta^{\hat i}}{dr}\nonumber\\
    &&-\frac{1}{rB^2}\left(\frac{NN'}{N^2-E^2}-\frac{B'}{B}\right)\eta^{\hat i}=0\ .
\end{eqnarray}
If the IC1 (\ref{IC1}) is applied to solve the equation of the angular geodesic deviation (\ref{geo-dev-ang-dif}) in the Schwarzschild spacetime ($\epsilon=a_1=b_1=0$), the equation can be solved analytically that gives the solution 
\begin{eqnarray}
    \eta^{\hat i}=\frac{r}{b}\ .
\end{eqnarray}
To determine the effect of the spacetime parameters, we solve the differential equation (\ref{geo-dev-ang-dif}) numerically and demonstrate the radial dependence of the angular component of the geodesic deviation vector in Fig. \ref{fig-eta-i-ic1} and \ref{fig-eta-i-ic2} for the IC1 and IC2, respectively.
If we consider the effect of the spacetime parameter $\epsilon$, one can see from the left panel of Fig. \ref{fig-eta-i-ic1} that the angular geodesic deviation vector always decreases and outside the horizon of the black hole it behaves almost the same for any value of spacetime parameters. Moreover, unlike the case in the Schwarzschild black hole, the particle falling from $r=b$ to the black hole does not shrink to zero size at radial coordinate by maintaining the finite size at $R^{\rm stop}$, instead of shrinking to zero size at the origin of the radial coordinate. With increasing the value of the parameter $\epsilon$ this finite minimum shrinking size increases. On the other hand, from the right panel of Fig. \ref{fig-eta-i-ic1} one can see that the effect of the parameter $a_1$ is opposite to the one of $\epsilon$, as with increasing the value of $a_1$, the minimum finite size of the particle decreases and at some point it even becomes zero.

In the case of the IC2, the angular component of the geodesic deviation vector behaves differently. One can see from Fig. \ref{fig-eta-i-ic2} that the behaviour of $\eta^{\hat{i}}$ is essentially the same outside the event horizon of the spacetime for various values of the spacetime parameters. It can be seen in Fig. \ref{fig-eta-i-ic2} that during the infall from $r=b$ to the black hole, $\eta^{\hat{i}}$ initially increases up to its maximum at about $r=b/2$ and decreases continuously until it reaches the black hole horizon. Inside the event horizon, the behaviour of the angular geodesic deviation vector is almost similar to the case of the IC1.

\section{Conclusion}\label{sec-conc}

In the current paper, we examined the tidal forces produced in the curved spacetime of a spherically symmetric static parametrized black hole. Our calculations reveal that unlike the behavior observed in the Schwarzschild black hole case where the radial tidal force (angular tidal force) continuously stretches (compresses) and diverges at the spacetime's center depending on the specific values of the spacetime parameters, in the parametrized black hole spacetime the radial and angular components of the tidal forces can be both positive (stretching) and negative (compressing) values inside the black hole horizon. Therefore, there is a point in the spacetime of the parametrized black hole where the specific component of the tidal force vanishes and beyond its behaviour changes from stretching (compressing) to compressing (stretching).

Additionally, we derived the equations governing geodesic deviation for particles in radially free-fall and proceeded to solve them through numerical calculations. Our analysis demonstrated that the spacetime parameters $\epsilon$ and $a_1$ yield significant influences on the magnitudes of the physical quantities related to tidal effects. It is well known that in Schwarzschild spacetime, a freely falling particle experiences consistent stretching in the radial direction and compression in the angular direction during its trajectory toward the spacetime center. In the parametrized black hole spacetime on account of the parameter $\epsilon$, the particle can only be stretched (compressed) to a finite size in the radial direction (in angular direction). On the other hand, the spacetime parameter $a_1$ attempts to counteract the influence of $\epsilon$ by contributing the stretching (compressing) nature of the radial (angular) tidal forces.

This study advances our knowledge of the complex interplay between spacetime parameters, tidal forces, and the behavior of test particles in curved spacetime produced by a black hole's gravitational field. 

As future prospects, one can develop astrophysical applications of tidal forces around black holes in order to get constraints on the parameters of RZ parametrization. In the comparison of the motion of close S stars precisely measured by the GRAVITY consortium in the close environment of a supermassive black hole Sgr A* one can get an estimation of tidal forces and get constraints on additional black hole parameters. Tidal effects, driven by varying gravitational forces, are pivotal in deciphering the enigmatic world of black holes. Recent research has illuminated their paramount importance, showcasing their influence on binary systems, spinning black holes, and accretion disks. Tidal forces sculpt the dynamics of these cosmic entities, inducing precessions, perturbations, and resonances, ultimately shaping our understanding of black hole physics and astrophysics. As we delve into this intricate cosmic interplay, it becomes evident that tidal effects hold the key to unraveling the mysteries of the universe's most enigmatic celestial objects, enriching our comprehension of the cosmos \cite{1997ApJ480G,101093pasj64,ABBAS2023101967,2013PhRvDj4010C,2014PhRvDY,2023PhRvD107l4039M,Yang:2017aht,Randall:2018qna,Bonga:2019ycj,Gupta:2019unn,Gupta:2021cno,Camilloni:2023rra,2010PhRvD81b4029P}. The study of tidal effects around astrophysics black holes provides insights into the fundamental nature of gravity, the behavior of matter under extreme conditions, and the astrophysical processes occurring in the vicinity of the gravitational compact objects.

\section*{Acknowledgements}

The authors thank the anonymous referee for his/her helpful comments that improved the quality of the manuscript. BT acknowledges the support of program ``CZ.02.2.69/0.0/0.0/18-053/0017871: Podpora mezin\'{a}-rodn\'{i} mobility na Slezsk\'{e} univerzit\v{e} v Opav\v{e}" at the Institute of Physics, Silesian University in Opava. Authors acknowledge the support of the Ministry of Innovative Development of the Republic of Uzbekistan Grants No.~F-FA-2021-510, F-FA-2021-432 and MRB-2021-527.

\bibliography{references}

\end{document}